# Terrestrial Life in Light of the Copernican Principle

Ian von Hegner
Future Foundation Assoc.
Egedal 21
DK-2690 Karlslunde

**Abstract** Although many solar systems have been discovered, only one example of life is known. Thus, terrestrial life represents merely one data point. Consequently, extrapolating from terrestrial life to life elsewhere in the galaxy and beyond is often seen as a limitation in the search for different forms of life. Essentially, attempting to extrapolate from terrestrial life to life elsewhere implies that terrestrial life is representative of all life, reflecting a geocentric viewpoint. However, in accordance with the Copernican principle, the opposite holds true. Asserting that terrestrial life must differ from other forms of life in the universe is, in fact, the geocentric viewpoint. For if life elsewhere is not like terrestrial life, then it is *ipso facto* different life; more precisely, if terrestrial life does not represent general life, then that life must represent special life, which the principle states it is not. This study employs the Copernican principle as a probability assessment, addressing critiques rooted in the implicit assumption of the existence of different extraterrestrial forms of life. If various fundamental forms of life indeed exist, then differences in the probabilities of their emergence can be expected, forming a probability scale. This holds significance because it not only allows for insights into the characteristics of the majority of life elsewhere but also facilitates the establishment of boundaries for categories of life as we do not know it. Thus, the Copernican-Darwinian principle provides a valuable tool for astrobiology and the search for life in the galaxy and beyond.

**Keywords:** astrobiology; Darwinian principle; extraterrestrial life; cosmological principle.

**1. Introduction**

In 1543 Copernicus changed the millennia-old geocentric model by in a geometrical sense removing the Earth and hence its life from the centre of the universe and replacing it with the sun. Hence, in the heliocentric model, the Earth lacks a privileged position in the universe and is now recognised as one of eight planets orbiting the Sun in an elliptical path, with the Sun occupying one focal point. An extension of this argument of a moving Earth has subsequently continued to remove the notion of a privileged position.

    Therefore, the Sun, positioned in the Orion arm of the Milky Way, is not privileged but is classified as a G-type main-sequence star [Morgan et al., 1952; van de Hulst et al., 1954], residing far from the galactic centre. The Milky Way forms part of the Local Group, [Hubble, 1936], which in turn is a constituent of the Virgo Supercluster [de Vaucouleurs, 1953; 1958]. This supercluster is integrated into the larger Laniakea Supercluster [Tully et al., 2014]. Indeed, the universe itself may be only one of many within a multiverse.

    This development, codified by the Copernican principle, has significantly influenced astronomical and cosmological discourse. It is posited here that this principle can also inform our approach to life beyond the Earth.

    Astrobiology is the multidisciplinary field that investigates the deterministic conditions and contingent events with which life arises, distributes and evolves in the universe [von Hegner, 2021]. Thus, while not technically focused on the search for extraterrestrial life, astrobiology contemplates the profound question of life elsewhere in the galaxy and beyond, including the status of terrestrial life in this context.

    While this solar system may harbour only one inhabited world, this does not prevent the potential for life to flourish elsewhere. With an estimated 100 to 400 billion stars in this galaxy alone [Williams, 2008], with virtually each now known to likely be accompanied by planets (and likely moons as well), and over 2 trillion galaxies in the observable universe alone [Conselice et al., 2016], the room for potential extraterrestrial life is indeed considerable.

    However, despite the abundance of known solar systems, the existence of life beyond the Earth remains unconfirmed, keeping the discourse within a geocentric framework.

Thus, so far only one data point is known for life, terrestrial life, meaning that at present all discussion of the nature of extraterrestrial life, apart from pure speculation, can only occur from a geocentric position, i.e. to extrapolate from terrestrial life to life elsewhere. While using only one data point to draw conclusions is a commonly criticized approach, the criticism, as will be discussed in this work, is not as strong as usually assumed when incorporating the Copernican principle. This approach allows for reasoned discussion on the nature of extraterrestrial life, even in the absence of empirical data.

Hence, the search for life elsewhere in the galaxy and beyond involves many unknown unknowns. By employing probability theory we however restrict ourselves to known unknowns. Thus, boundaries can be placed upon an unknown by focusing on a known, terrestrial life. Thus, despite the term 'geocentric' inherently referring to a single world, it can nevertheless provide insights into life on a universal scale, as elaborated upon in the subsequent discussion.

**2. Discussion**

In this paper, Section 3 elucidates how the Copernican principle can justify the extrapolation from terrestrial life to life elsewhere. Section 3.1 delves into the asymmetry inherent between the potential forms of life and their abundance. Section 3.2 provides and analyses calculations supporting the notion that terrestrial life likely belongs to the general form of life in the galaxy and beyond. Section 3.3 explores the significance of extrapolation being grounded in chemical and biological evolution. Moving forward, Section 4 delineates the content of the Earth's one data point for life, arguing that the bacterial/archaeal form represent the general form of life. Section 4.1 explores what can be understood by other forms of life and the evolutionary mechanisms governing life. Additionally, Section 4.2 sheds light on the relationship between intelligent life and microbial life. Finally, Section 5 encapsulates the findings of this study and discusses their implications for applying the Copernican principle to gain insights into the majority of life in the galaxy and beyond.

**3. The Copernican principle**

Terrestrial life so far represents only one data point. It represents, as the phrase goes, life 'as we know it.' It is, therefore, plausible there are examples of life elsewhere in the universe that are very different.

Thus, extrapolating from terrestrial life to potential life elsewhere in the galaxy and beyond is commonly perceived as imposing limitations on the search for life elsewhere. In other words, to extrapolate from terrestrial life to life elsewhere is to say that terrestrial life represent universal life, thus embodying a geocentric viewpoint.

However, according to the Copernican principle, the inverse holds true. Arguing that terrestrial life ought to differ from other potential life forms in the universe is in fact the geocentric viewpoint.

The Copernican principle states that 'the Earth is not in a central, specially favored position .. the Earth is in a typical position' [Bondi, 1952]. Therefore, observations made from the Earth are merely indicative of observations made from an average vantage point in the universe. While Copernicus himself did not formulate this principle bearing his name, this was done by Bondi (1952), the principle is a retrospective extension of the ongoing deconstruction of any privileged status accorded to the Earth—a generalization that has proven notably fruitful. The Copernican principle pertains to the cosmological principle, according to which 'the universe presents the same aspect from every point except for local irregularities' [Milne, 1935; Bondi, 1952].

The Copernican principle thus has its application in an astronomical and cosmological connection. However, the fact remains that like everything, life too forms part of the universe, and the application of the Copernican principle to life is thus considered a natural extension in this work.

Hence, the Copernican principle asserts that the Earth occupies no special position, and therein lays the central issue. If terrestrial life does not represent general life, it *ipso facto* represents special life. The situation is inversed, so to speak.

Thus, that terrestrial life is not general contradicts the principle. If the Copernican principle is taken seriously, then the critique of extrapolating from terrestrial life to life elsewhere is grounded in the assumption that terrestrial life is different from life forms elsewhere—a presumption directly contradicted by the principle, which posits that the Earth do not have a privileged status in the cosmos.

As terrestrial life so far constitutes only one data point, concerns emerge regarding extrapolation. The term extrapolation refers to the use of a data set to predict values in relation to this data. It can be broadly defined as 'an inference made about a system's behavior in a new range of variables, from experience in an old (familiar) range' [Matalas and Bier, 1999]. Thus, extrapolation refers to predicting values beyond a specific range of known data points. It is related to interpolation, which can be defined as 'a method for estimating the value of a function between two known variables' [Rukundo and Schmidt, 2018], or predicting values that are inside a range of existing data points, but extrapolation is subject to greater uncertainty.

Given the singular nature of this data point, extrapolating from terrestrial life is considered to introduce considerable uncertainty, thus constraining the search for extraterrestrial life in the galaxy and beyond.

However, the application of the Copernican principle in this context shows that two implicit assumptions shape this criticism, meaning the criticism with or without the principle has fundamental differences for the discussion.

Firstly, the presumption of extraterrestrial life's existence, which, while trivial, forms a foundational premise for further discussion. If life is not assumed to exist elsewhere, then further discussion becomes redundant.

As per the second implicit assumption, extraterrestrial life must be different from terrestrial life. Unlike the first assumption, this one is non-trivial because if life elsewhere is not presumed to be different, then the critique becomes redundant. Therefore, a critique of extrapolating from terrestrial life is meaningful only if the existence of different life is implicitly assumed; extrapolation from terrestrial life in this connection is *ipso facto* justifiable if life is the same everywhere in the universe.

Thus, these implicit assumptions bring the matter straight to the core of the Copernican principle even if the extrapolation with the one data point in itself is not mathematically justifiable. As explained, this posits a fundamental premise that the Earth, along with its life forms, lacks a privileged position in the galaxy and beyond. If terrestrial life does not hold a privileged or special status compared to life forms that are prevalent throughout the universe, then it is reasonable, based on this rationale, to employ an extrapolation of terrestrial life as a foundation for discussing life elsewhere in the galaxy and beyond.

*3.1. The form and amount of life*

Given that only one data point has been established so far, extrapolating from terrestrial life to potential life elsewhere can be viewed as a geocentric approach. Nevertheless, asserting that life elsewhere in the galaxy and beyond differs from terrestrial life is, as explicated, in fact the geocentric viewpoint.

Extrapolation involves inherent uncertainty, as it unlike interpolation entails predicting values that are outside of a range of known data points, making inferences based on an assumed correspondence that may not necessarily exist.

Despite the current absence of empirical evidence regarding the existence of life elsewhere in the galaxy and beyond, the Copernican principle can still offer insights into the discussion through probabilistic evaluations. While this doesn't affirm the existence of life elsewhere, it provides a framework for meaningful discourse.

Thus, the crux of the matter lies in recognising that if life elsewhere diverges from terrestrial life, then it is *ipso facto* a different form of life. Put succinctly, if terrestrial life is not general life, it is *ipso facto* special life.

However, categorising some forms of life as 'general' and others as 'special' appears to stem from an anthropocentric perspective. Such classification prompts the question of what criteria delineate general versus special life forms.

A Darwinian principle can be posited, asserting that no single species holds a privileged position in nature [von Hegner, 2019]. According to this principle, *Homo sapiens* do not enjoy any privileged status, mammals do not enjoy any privileged status, animals do not enjoy any privileged status, and so forth. Thus, there is no such thing as lower and higher life in evolution, there is only reproductive success.

The cosmological principle allows irregularities locally, which can be interpreted as the possibility of variations of forms of life. This suggests that in a galaxy hosting inhabited worlds with diverse forms of life, an inherent asymmetry exists in the distribution of these life forms. Despite the absence of any form of life holding a special status, variations in the abundance of different life forms across worlds are to be expected. Each form of life remains without special status, yet their distribution among different worlds varies in quantity.

When referring to 'different life' here, we extend beyond the diversity observed on the Earth. It is noteworthy that all known life on the Earth shares a common ancestry, with origins traced closest to the bacterial domain [Madigan et al., 2022]. Therefore, when we mention terrestrial life or the form of life, we specifically mean the bacterial/archaeal form, elaborated upon in section 4. The diversity in life emerges from the evolutionary process of speciation, where ancestral groups diverge into numerous subgroups over time, each adapting to distinct ecological niches, thereby accumulating differences. Thus, when asserting that life differs between inhabited worlds, it implies fundamental distinctions, such as variations in genetics and biochemistry, or even the potential existence of life devoid of conventional genetics and biochemistry, thus sidestepping anthropocentric perspectives.

In a galaxy housing inhabited worlds with diverse life forms, this diversity encompasses different classifications of life. If various forms of life exist, the question emerges: how much of one class of life exists compared to others? Due to deterministic and stochastic factors, these diverse forms of life cannot be equally distributed; there must be an asymmetry in their abundance.

Consequently, if distinct fundamental forms of life exist, they would also manifest varied pathways for their emergence, resulting in differences in the probabilities of their occurrence. While specific figures for these probabilities remain unknown, the concept holds true in principle.

Hence, a probability scale emerges wherein certain life forms exhibit a higher improbability of emergence compared to others, creating an asymmetry in the probability of their occurrence. Consequently, some life forms emerge more readily, while the emergence of others is so cumbersome that it faces a higher improbability.

Although these diverse life forms likely emerge due to the same known physical and chemical laws, it is the case that even if there are undiscovered laws of nature affecting the probabilities of the emergence of some of these forms, even then will there be a difference between all these forms of life in terms of probability.

Consequently, variations in the probability of different life forms emerging also translate into statistical differences in their prevalence between inhabited worlds and the abundance of forms of life in the galaxy and beyond. As a result, some forms of life will be rarer and exist on fewer worlds than others, introducing statistical inequalities among them. Thus, there is not equality between the different forms of life, meaning that life through this asymmetry between forms of life can appear special by existing on fewer worlds and in smaller quantities in the galaxy than other forms of life.

Moreover, stochastic events further influence the prevalence of different life forms across inhabited worlds. Thus, even if different forms of life were to emerge with identical probabilities, purely statistical events would ensure variations in their quantities. Thus, contingent events would ensure some forms of life being more prevalent than others in the group of inhabited worlds.

In this context, 'general' refers to the most prevalent form of life in the galaxy and beyond, characterised by its widespread existence among inhabited worlds. Conversely, it is possible for life to appear 'special' since it denotes the least prevalent form when compared to another more prevalent form. However, as discussed, different probabilities will be assumed for different forms of life, and these will determine the amount of different life.

Of course, the origin of life on each world are independent events, i.e. what happens on a world A is independent of what happens on another world B at the other end of a galaxy. However, although life emerges independently of each other on different worlds, the form of life for each of the mentioned probabilities can still be the same. Thus, a certain form of life, e.g. the bacterial/archaeal form, may only emerge that way, even on different worlds.

Thus, life being placed in the same class if possessing the same form, yet having emerged independently of each other on separate worlds, is in principle no different than e.g. two G-type main-sequence stars at opposite ends of the galaxy being treated the same way in astrophysics. These stars have emerged the same way and follows the same phases, thus, they are placed within the same type where an understanding of the sun makes it possible to understand another type of G-type main-sequence star as well.

Thus, although life emerges independently of each other on different worlds, if it has the same form, it will here be considered as belonging to the same class of life.

More than one form of life may potentially emerge on the same world, however, competition for resources and space often results in one form outcompeting the others. This dynamic may lead to one form becoming dominant, while the others persist as minorities, inevitably resulting in differences in their abundance within a given world.

*3.2. The probability of forms*

As its basic postulate, the Copernican principle states that the Earth, and thus its life, does not hold a privileged position in the galaxy and beyond. However, is this justifiable?

The crux of the criticism against extrapolating from terrestrial life lies in the implicit assumption that terrestrial life holds a special or privileged status, indicative of rare forms of life. If all life resembles terrestrial life, then the Copernican principle stands in its purest form. Alternatively, if certain irregularities are permitted in terms of differences in the form and abundance of life, as allowed by the cosmological principle, then we adhere to a quasi-Copernican principle known as the mediocrity principle. Under this principle, we assume mediocrity rather than initially presuming any phenomenon to be special or privileged.

Consequently, if an object is randomly selected from multiple categories, it is more likely to belong to the most numerous category than any of the less numerous ones. While it is not a given that terrestrial life cannot fall into the less numerous categories, it is more likely for it to belong to the most numerous category.

Thus, it comes down to probability theory, which can be illustrated by the following relatively straight forward example. It can be assumed for the sake of argument that life can emerge in four fundamentally different forms according to a probability scale: $N_{lifeforms} = P_1 > P_2 > P_3 > P_4$.

One estimate suggests that there could be at least 300 million potentially habitable worlds within this galaxy [Bryson et al., 2021]. Therefore, for the purposes of discussion, let us assume that life has emerged on 100 million worlds within the galaxy.

Among the various life forms, life form 1 prevails as the dominant one, inhabiting 80 million worlds. Following this, life form 2 inhabits 10 million worlds, while life form 3 inhabits 6 million worlds, and life form 4 inhabits 4 million worlds Thus, there is a total of $8 \times 10^7 + 1 \times 10^7 + 6 \times 10^6 + 4 \times 10^6 = 1 \times 10^8$ inhabited worlds in the galaxy, where one world out of $1 \times 10^8$ can be drawn by $^{1 \times 10^8}C_1$ ways. Hence,

$$n(S) = {}^{1 \times 10^8}C_1 = 1 \times 10^8 \qquad (1)$$

The probability of randomly drawing life form 1 is as follows: Let A be the event of getting life form 1 on a world. There are life form 1 on $8 \times 10^7$ worlds in the galaxy, and 1 world out of $8 \times 10^7$ worlds can be drawn by $^{8 \times 10^7}C_1$ ways. Thus,

$$n(A) = {}^{8 \times 10^7}C_1 = 8 \times 10^7 \qquad (2)$$

Therefore, the probability of drawing life form 1 on a world is as follows:

$$P(A) = \frac{n(A)}{n(S)} = \frac{8 \times 10^7}{1 \times 10^8} = 0.80 \qquad (3)$$

For the other life forms, the same method can be followed:

The probability of randomly drawing life form 2:

$$P(B) = \frac{n(B)}{n(S)} = \frac{1 \times 10^7}{1 \times 10^8} = 0.10 \qquad (4)$$

The probability of randomly drawing life form 3:

$$P(C) = \frac{n(C)}{n(S)} = \frac{6 \times 10^6}{1 \times 10^8} = 0.06 \tag{5}$$

The probability of randomly drawing life form 4:

$$P(D) = \frac{n(D)}{n(S)} = \frac{4 \times 10^6}{1 \times 10^8} = 0.04 \tag{6}$$

There is also the following illustrative example. The probability of not randomly drawing life form 1: Let E denote the event of not getting life form 1 on a world. There are $1 \times 10^7 + 6 \times 10^6 + 4 \times 10^6 = 2 \times 10^7$ non life form 1 on $8 \times 10^7$ worlds, and 1 non life form 1 world out of $2 \times 10^7$ non life form 1 worlds can be drawn by $^{2 \times 10^7}C_1$ ways. Hence

$$n(E) = {}^{2 \times 10^7}C_1 = 2 \times 10^7 \tag{7}$$

Therefore, the probability of not randomly drawing life form 1 is:

$$P(E) = \frac{n(E)}{n(S)} = \frac{2 \times 10^7}{1 \times 10^8} = 0.20 \tag{8}$$

For the other life forms, the same method can be followed:

The probability of not randomly drawing life form 2:

$$P(F) = \frac{n(F)}{n(S)} = \frac{9 \times 10^7}{1 \times 10^8} = 0.90 \tag{9}$$

The probability not randomly drawing life form 3:

$$P(G) = \frac{n(G)}{n(S)} = \frac{9.4 \times 10^7}{1 \times 10^8} = 0.94 \tag{10}$$

The probability of not randomly drawing life form 4:

$$P(H) = \frac{n(H)}{n(S)} = \frac{9.6 \times 10^7}{1 \times 10^8} = 0.96 \tag{11}$$

Hence, considering terrestrial life as more likely to belong to the most numerous category rather than any of the less numerous categories appear statistically sound. From a statistical standpoint, the probability that terrestrial life among 100 million inhabited worlds falls within one of the 80 million worlds is notably higher. Consequently, the Copernican principle becomes a statistical argument—a measure of probability concerning both the form and abundance of life, comparing terrestrial life to life elsewhere.

Although the numerical values provided were arbitrary, as we lack empirical knowledge of life beyond the Earth, the principle still applies, regardless of differences in actual conditions or the actual abundance of forms of life. This also applies if the different forms of origin of life are not probable events, as these still have different probabilities compared to each other.

This still does not make any of the forms of life lower or higher than others, it is merely a statistical difference between different inhabited worlds and the difference in their abundance in the galaxy and beyond given by the determinism of physical and chemical processes as well as the contingency of biological events.

Thus, the Darwinian principle integrates into the Copernican principle, which underscores the absence of privileged species in any real vantage point, emphasising instead statistical differences in the occurrence of forms of life.

However, it is not a given that terrestrial life cannot belong to one of the less numerous categories. Among 100 million inhabited worlds, terrestrial life may potentially be among the 4 million worlds, rather than the 80 million worlds. Thus, on the probability scale, there exists life whose emergence stems from less likely events, placing it in the minority of the universe's forms of life in terms of probability and numbers across numerous worlds. However, does terrestrial life belong to it?

The bacterial/archaeal form appears to be so rudimentary and elementary that it can be argued to be the one that emerges through the most likely events, which will be discussed in greater detail in section 4.

Yet, terrestrial life could potentially belong to the rare life. It may be special, there is no empirical data to support either option yet. Consequently, criticism regarding extrapolation from terrestrial life in this regard is meaningful.

Nevertheless, if compelled to make a wager, a prudent approach would be to bet on terrestrial life belonging to the most numerous category, i.e. terrestrial life is not privileged. It is therefore more justifiable to assume that terrestrial life is not special. Along this line of reasoning, criticisms regarding extrapolation from terrestrial life lose their potency. Ultimately, the decision rests on whether to embrace probability theory as a guiding principle in astrobiological research.

*3.3. Chemical and Biological evolution*

Extrapolation arguments may be strong or weak depending on how far one venture from known data points, potentially leading to less precise predictions.

Since the majority of life elsewhere has the form of terrestrial life, extrapolation will *ipso facto* be strong from its grounding in knowledge of chemical evolution, i.e. the natural assembly of complex molecules into living entities; and biological evolution, i.e. the natural ability for these living entities to use materials in the environment to generate variable copies of themselves.

Biological evolution provides profound insights into life's mechanisms, its propagation, adaptation to changing environments, and the dynamics of species emergence and extinction. Meanwhile, chemical evolution provides an increasing knowledge on the origin of life on this planet, and through this offers insights into the majority of life within the galaxy and beyond.

The significance of terrestrial life as a representation of general life cannot be overstated. It suggests that much can be inferred about extraterrestrial life based on the understanding of life on the Earth, even though life has not yet been located elsewhere.

However, it is important to realize that since terrestrial life belongs to general life, this implies that there can be special life, i.e. different life, of which, on the basis of terrestrial life, nothing may be said. Thus, since a minority has forms of life fundamentally different from the form of terrestrial life, then extrapolation will be weak, as it may not be possible to ground it in knowledge of terrestrial life. Until research into chemical evolution may unveil insights into these potential other forms of life, which will also make it possible to formulate numbers on the probability scale for life's mutual probabilities, criticism of extrapolation from a solely terrestrial perspective is justifiable, as this geocentric approach may hinder the recognition of different life forms, even if they are present in front of us.

Yet, since this life belongs to the minority in the galaxy and beyond, knowledge of the majority of life can still be obtained since terrestrial life belongs to the life which emerges most easily and exists on the largest number of worlds. Thus, it does not impair the possibility to work with the majority of life elsewhere, and the advantage will therefore far outweigh the disadvantage of this approach. The Copernican principle can thus be meaningfully applied even though life may exist whose form escapes the current understanding of life. Thus, it is still a sound approach.

It is also conceivable that is terrestrial life that holds a special status, and an extrapolation of this life may thus potentially prevent us in seeing the different, yet more general life elsewhere. However, as scrutinised through the lens of the Copernican principle, it is more likely that terrestrial life is not privileged, meaning it is still a more reasonable bet that terrestrial life holds the general status.

The extrapolation will be weak in terms of specifying what types of species there can be on other worlds besides the bacterial/archaeal form, which is due to the way evolution operates, i.e. specifying the specific ecosystems on other worlds.

Thus, evolution is about local adaptation to immediate environments. These environments are changing effectively on random vectors through geological time. This means that even if there is determinism of local adaptation for any given moment, since it is determinism of local adaptation to fluctuating environments that are effectively stochastic through geological time, it will be a stochastic pattern for the history of life [Gould, 2002].

Thus, on an inhabited world the history of life will be unpredictably contingent since that history follows a stochastically changing vector of environmental circumstances through geological time. But the extrapolation will be strong with regard to the general overview, where the bacterial/archaeal form will be prevalent.

Contrary to the view that one example of life represents the geocentric limitation, this one example do not represent a limitation in our knowledge of life in the universe. Thus, even in the current absence of data on life elsewhere, it is the case that since terrestrial life likely belongs to the majority, we possess an extensive knowledge regarding the nature of most life potentially existing in the galaxy and beyond. As our understanding of chemical and biological evolution continues to increase and refine, it is according to the Copernican principle justifiable to utilize that knowledge of terrestrial life as search criteria for life elsewhere in the galaxy and beyond.

**4. The data point content**

It has been reiterated that terrestrial life only represents one example, a single data point from which we extrapolate. However, a closer examination of life on the Earth reveals a vast diversity of organisms, many of which exhibit morphological differences so significant that one might not readily discern their relatedness.

Despite this apparent diversity, all known life on the Earth shares a common genetic and biochemical structure, pointing to a common origin [Madigan et al., 2022]. This commonality underscores there is indeed only one common data point for life on the Earth, rooted within the bacterial domain from which all terrestrial life has diverged into an ever-branching bush.

When discussing the Earth as one data point, its form of life, we are effectively referring to microbial life, such as bacteria and archaea. Indeed, the Earth is fundamentally a microbial world, with multicellular life emerging relatively late in its history. Thus, if all plant and animal life were somehow eradicated, life on the Earth would remain essentially unchanged, but the eradication of microbial life would quickly lead to a lifeless world. Thus, this is the age of microbial life, which occupies virtually every available niche on the Earth, which it will continue to do until the phases of the sun towards the red giant stage put an end to this, the oldest terrestrial life.

However, while this holds true for the Earth, it doesn't necessarily mean it holds true on other inhabited worlds. The crucial question becomes whether the emergence of the bacterial/archaeal form represents the least improbable path.

The Darwinian principle states that no single species holds a privileged position in nature. Thus, *Homo sapiens* do not hold a privileged position, mammals do not hold a privileged position, indeed, multicellular life does not hold a privileged position in nature. However, when considering terrestrial life, this principle encounters its limits at the level of bacteria and archaea. These organisms cannot for this one world meaningfully be contrasted against anything, because what can they truly be contrasted against? It cannot be stated whether they are privileged or general, they simply are!

However, when the Darwinian principle integrates into the Copernican principle, a meaningful discourse emerges within a galactic context. Here, the focus shifts back towards the mediocrity principle, where no initial assumption of specialness is made.

Thus, that the bacterial/archaeal form belongs to the most numerous category may stem from that life cannot be simpler than the most rudimentary bacterial or archaeal form imaginable; this structural form seems to emerge from the simplest conceivable complexity of life. As such, it is plausible that the majority, if not all, of life elsewhere must necessarily emerge or be instantiated in a form akin to the simplest functional bacteria or archaea possible.

Hence, one could posit that the bacterial/archaeal form is identical to the universal standard for emerging life. However, the state of affairs is not quite as straightforward.

Thus, if we identify the universal standard for life with the bacterial/archaeal forms nature, then it appears we are identifying it with certain properties of the bacterial/archaeal form. If so, a dilemma appears: Is the bacterial/archaeal form the universal standard because it has those properties, or are those properties the universal standard because the bacterial/archaeal form has them?

Or alternatively: Did chemical evolution discover the bacterial/archaeal form because it is the universal standard, or is the bacterial/archaeal form the universal standard because chemical evolution discovered it?

These issues currently pose significant challenges, and it is therefore not a given that the bacterial/archaeal form cannot belong to the less numerous categories. However, when the Darwinian principle is generalized into the Copernican principle, it becomes more likely that this form indeed fall within the most numerous category, i.e. the general form. This suggests that certain forms of life may emerge with a higher probability compared to others, as discussed with the probability scale.

Life as we know it is a planetary phenomenon. However, speculation persists regarding the existence of space life, exemplified by fictional entities like the black cloud organism proposed by Hoyle (1957), albeit this may lean more towards a Lamarckian organism rather than a Darwinian organism, some scientific support exists for their plausibility [Tsytovich et al., 2007]. Yet, even if such or other kinds of space life were to exist, differences in probability and quantity of different life forms would still persist. Thus, would the emergence of such space life prove more cumbersome than the emergence of the bacterial/archaeal form?

The bacterial/archaeal form appears to be so rudimentary and elementary that it likely emerges with greater ease compared to other life forms.

Indeed, it is not beyond the realm of possibility that life arose in multiple waves in this form on the Earth. Thus, it has been proposed that bacteria and archaea represent two distinct origins of life on this planet [von Hegner, 2020]. Their cellular similarities stem partly from inherent constraints dictating their formation only in this way, allowing for some degree of variation, and partly from gradual gene exchange through lateral gene transfer.

Hence, it is plausible that these organisms belong to the most numerous category of life and lack any privileged status in the galaxy and beyond.

Another rationale for the potential inclusion of the bacterial/archaeal form within the most numerous category stems from its inherent simplicity. Thus, it is not merely the ease of its emergence that is decisive, but also the rapidity and range of its operation. This form of life exhibits distinct traits concerning a world, characterised as 'information explosive' [von Hegner, 2021]. Upon emergence, it swiftly proliferates and disseminates, eventually occupying every available niche within a world.

Under certain conditions, such life forms can even 'leak' from a world, as on the Earth where it may be propelled into space by external impacts [von Hegner, 2020], or potentially on the Saturnian moon Enceladus through internal plumes [Porco et al., 2017], if such life exists there. Hence, this form of life plays a dual role, planetary in its capacity to rapidly occupy and dominate a world, and interplanetary by potentially transferring to and colonising neighbouring worlds.

Therefore, its velocity and range serve as pivotal factors that could render it the most prevalent form of life in the galaxy and beyond.

*4.1. The other forms of life*

The assertion has been put forth that terrestrial life is more likely in occupying the most numerous category, thereby positioning it as the general form of life. However, this assertion prompts a critical inquiry: is it even possible to address categories of life as we do not know it, such as e.g. the example of 3 other forms of life? After all, it is an implicitly given condition that it is life as we do not know it.

However, since all these forms of life must necessarily be part of this universe, regardless of whether they emerge by laws of nature either known or not yet known, that objection entails a paradox. This can be analyzed in the following.

In space A, there is a set consisting of 'life as we do not know it', and in space B, there is a set consisting of things that are not 'life as we do not know it'. Thus, 'life as we do not know it' goes in space A, and 'life as we know it' goes in space B.

The question now emerges, where does space A belong? It is evidently not 'life as we do not know it'. This means it must go in space B.

However, this gives rise to a problem, because space B is defined to only contain things that are not 'life as we do not know it'. This means that the contents of space A cannot go in space B.

Since space A, or the set, is not, in itself, 'life as we do not know it', but, by its definition, contains 'life as we do not know it', we cannot effortlessly categorise where the set belongs.

This paradox is, strictly speaking, an illustration of the notion that there is no such entity as life as we do not know it, or, put differently, that the condition itself is inconsistent. In essence, it highlights the challenge in comprehending the concept of life as we do not know it, not solely because we lack empirical encounters with this 'dark biomatter', but also because it must necessarily be part of this universe and we thus are compelled to acknowledge that we know it.

This serves a significant point, as life as we do not know it is not the same as saying there is life we cannot explain.

Thus, 'n is a bachelor' means 'n is an unmarried adult male'. As a consequence, 'n is a bachelor but n is not an unmarried adult male' is false on its face; and 'if n is a bachelor, n is an unmarried adult male' and 'if n is an unmarried adult male, n is a bachelor' are tautologies.

By contrast, 'life as we do not know it but it is life that we can explain' is not false on its face; and 'life as we do not know it is life that we cannot explain' and 'life that we cannot explain is life as we do not know it' are not mere tautologies.

This is sufficient to dispose of the claim that 'life as we do not know it, and 'life that we cannot explain' has the same meaning.

The necessity of life being part of this universe, subject to its laws of nature, leads us to a related yet distinct inquiry. Life as we know it is an evolutionary phenomenon, with the Earth teeming with Darwinian organisms. However, it might be opinionated that this is merely due to terrestrial life being guided by evolution as we know it.

This begs the question: Is this evolution universal? Could the other forms of life differ not only in their origins but also in their operational dynamics, potentially influencing their prevalence throughout the galaxy and beyond?

To delve deeper into this inquiry, it is essential to clarify our understanding of evolution. Evolution by natural selection is fundamentally a theory about interactors; it is not about replicators. It is about individual organisms that interact with their local environment, where natural selection means differential reproductive success conferred by the fitness of possessing certain beneficial properties. Thus, Darwinian interactors can be replicators such as genes in some circumstances, while individual organisms are the interactors in most of circumstances.

Thus, while life as we know it is a DNA or RNA based phenomenon, organisms elsewhere in the cosmos might conceivably possess hereditary material in forms divergent from the standard 4-base pair DNA and RNA molecules, such as DNA and RNA molecules consisting of 8-base pair [Hoshika et al., 2019], or even in a form other than genes. However, such deviations do not alter the fact that organisms with alternative genetic materials remain subject to the workings of evolution.

Thus, in one interpretation, evolutionary mechanisms are supervenient on the hereditary material, because while a given hereditary material specifies an evolutionary strategy, a given evolutionary strategy does not specify any finite set of hereditary material.

Thus, the question emerges: If life on different worlds exhibits difference at the hereditary material level but evolution acts the same at the organismal level, can it truly be considered fundamentally different in the universe? The answer hinges on the prioritisation of either level. However, if disparate forms of life are governed by evolution the same way at the organismal level, their differences may be inconsequential in the broader context of cosmic existence.

*4.2. Intelligent life*

Expanding the discourse to encompass life beyond the Earth naturally leads to considerations regarding the potential existence of intelligent life elsewhere in the galaxy and beyond. While our focus thus far has centred on the bacterial/archael forms of life, with multicellular organisms being comparatively recent additions, one might argue that the emergence of intelligent terrestrial life could confer a privileged status upon the Earth.

An occasionally espoused viewpoint posits that microbial life may be prevalent throughout the galaxy, yet intelligent life remains rare. However, this viewpoint is internally contradictory [von Hegner, 2022].

Thus, in microbiology, there is a growing recognition that microbial life can and indeed does exhibit characteristics of intelligence. Macromolecular networks can facilitate intelligent characteristics in microbial organisms such as decision-making, association and anticipation, self-awareness, robust adaptation, and problem-solving capabilities [Westerhoff et al., 2014].

Thus, the presence of microbial and protist forms of life on a world elsewhere in the galaxy does not preclude the existence of intelligent life. Consequently, the traditional conceptual dichotomy between microbial life and intelligent life is a misnomer that should be discarded. Indeed, viewing them as distinct categories hampers the search for intelligent life beyond this planet [von Hegner, 2022].

In this context, the crucial consideration becomes the emergence of the form of life, and it is the focus on the prevalence of inhabited worlds and the abundance of life therein that is important.

**5. Conclusion**

Since 1543, a geocentric approach has with great success no longer been used within astronomy and cosmology. However, while the geocentric approach has been abandoned there, it is in some ways the very same approach in light of the Copernican principle that, seemingly paradoxically, can give it a renaissance within astrobiology.

The Copernican principle is neither a law of nature nor a theorem, but is at present a heuristic, a working assumption derived from Copernicus' argument of a moving Earth, and therein lies its limitation. However, in this work the principle has been utilized as a measure for probability, as there may be irregularities in the form of different classes of potential life on different inhabited worlds and different amounts of life, thus, it will be mathematical in nature.

So although it is not yet known whether there is life elsewhere in the galaxy and beyond, the strength of the Copernican principle lies in its probability assessment. It allows, in light of the implicit assumptions, to frame the situation through the various inhabited worlds, and thus put some boundaries on what can be stated about life universally.

However, some issues persist. For instance, the numerical values used in the probability calculations was arbitrary, yet the underlying approach remains sound as it highlights the differences between potential forms of life. Yet, understanding how these potential differences correlate remains a challenge.

Thus, if terrestrial life remains the sole example in the cosmos, it would indeed hold a privileged status.

However, the discovery of life or remnants thereof on one of the other solar system bodies may not be decisive for whether it is privileged. If this life is similar to terrestrial life, it could merely suggest mechanisms such as lithopanspermia, wherein life is transferred between the different worlds [von Hegner, 2020]. Thus, the geocentric extrapolation will merely slip into a heliocentric extrapolation.

Alternatively, the conditions within this solar system may be conducive only to the emergence of a certain form of life. This could mean that a galaxy could be divided into many different regions, such that conditions within different star systems may result in regions conducive to specific forms of life while inhibiting others. This does not imply privilege for certain forms but rather underscores that different star systems provide different conditions for the emergence of life.

If life is discovered on one of the other solar system bodies that is distinct from terrestrial life, then it applies that the quantity of life on one world will surpass that of another, not necessarily due to differences in the probability of life's emergence, one being more cumbersome than the other, but rather as a result of contingent events shaping each world's evolutionary trajectory.

Thus, the mere identification of life elsewhere, marking the attainment of the next data point, will not suffice to conclusively determine whether the Earth holds a privileged status.

In scenarios where numerous inhabited worlds exist however, the relationship between general and special forms of life becomes a non-trivial question.

Thus, if the majority of extraterrestrial life is similar to terrestrial life in terms of emerging in the bacterial/archaeal form, the quasi-Copernican principle, then this supports the basic premise in this article, since the extrapolation of terrestrial life then will be meaningful.

In a similar vein, if all extraterrestrial life emerge in the bacterial/archaeal form, the Copernican principle in its purest form, then this supports the basic premise even more, namely that an extrapolation from chemical and biological evolution is justified.

While it remains possible that terrestrial life belongs to the minority of life, statistics suggest otherwise. Even in the absence of confirmed extraterrestrial life, statistical analysis indicates a higher probability of terrestrial life representing the majority.

Thus, despite the arbitrary nature of exact estimates in the calculations, the principle allows us to delineate broad boundaries regarding the nature of life in the universe.

It ultimately comes down to whether one wants to take probability theory seriously and apply it as a guiding principle in research.

Thus, the criticism against extrapolating from terrestrial life to life elsewhere is, within an astrobiology context, not as straight forward as otherwise thought, since it is precisely because the Copernican principle states that the Earth, and by extension, terrestrial life, is not privileged that makes it so useful. It thus offers a fruitful and robust tool for the search for life elsewhere in the galaxy and beyond.

The Earth has in recent centuries repeatedly been stripped of its privileged status in the universe, which has often been regarded by some as a diminution in self esteem. However, even though the Earth's life here too loses a presupposed privileged status, this can almost paradoxically be argued to be a source of increased self esteem. Thus, that terrestrial life likely belongs to general life has tremendous significance, as it is precisely this that means that much can be known about the majority of life elsewhere in the galaxy and beyond on the basis of knowledge of this life.

In the context of astrobiological research, terrestrial life is special in being near-universal, allowing it to make inferences about the majority of life elsewhere in the galaxy and beyond. Thus, while the term geocentric by its very meaning is limited to one world, it still inform nearly universally about life. This recognition does not confer terrestrial life the status of being the center of life in the universe, it is still not privileged. It merely underscores terrestrial life sharing its form with the majority of life in the universe.

Astrobiology is in many ways the offspring of the Copernican and Darwinian revolutions, the great thresholds in history which adjusted humankind's understanding of the cosmos and life. It is therefore appropriate that a Copernican-Darwinian principle can allow us to make informed statements regarding the distribution of life in the universe. Thus, just as the heliocentric model unlocked the study of other worlds, since the idea that other planets could be worlds like the Earth conceptually did not exist in the minds of natural philosophers before then, and just as evolutionary theory unlocked the possibility of studying life across past, future and remote places in the present day, the Copernican-Darwinian principle unlocks the possibility to work with potentially inhabited worlds elsewhere in the galaxy and beyond.